\documentclass[
superscriptaddress,
twocolumn,
amsmath,amssymb,
aps,
prb,
]{revtex4-2}

\usepackage{xcolor}
\usepackage{amsfonts}
\usepackage{epsfig}
\usepackage{psfrag}
\usepackage{mathrsfs}
\usepackage{esint}
\usepackage{pifont}
\usepackage{url}
\usepackage{nicefrac} % For comparison
\usepackage{stackengine}
\usepackage{comment}
\usepackage[T1]{fontenc}

\definecolor{melanzana}{RGB}{127,0,127}
\definecolor{verde_ale_chiaro}{RGB}{79, 163, 150}

  \definecolor{seashell}{rgb}{1.00, 0.96, 0.93}
  \definecolor{honeydew}{rgb}{0.94, 1.00, 0.94}
  \definecolor{mintcream}{rgb}{0.96, 1.00, 0.98}
  \definecolor{azure}{rgb}{0.94, 1.00, 1.00}
  \definecolor{aliceblue}{rgb}{0.94, 0.97, 1.00}
  \definecolor{lavender}{rgb}{0.90, 0.90, 0.98}
  \definecolor{lavenderblush}{rgb}{1.00, 0.94, 0.96}
  \definecolor{mistyrose}{rgb}{1.00, 0.89, 0.88}
  \definecolor{white}{rgb}{1.00, 1.00, 1.00}
  \definecolor{black}{rgb}{0.00, 0.00, 0.00}
  \definecolor{midnightblue}{rgb}{0.10, 0.10, 0.44}
  \definecolor{navy}{rgb}{0.00, 0.00, 0.50}
  \definecolor{navyblue}{rgb}{0.00, 0.00, 0.50}
  \definecolor{cornflowerblue}{rgb}{0.39, 0.58, 0.93}
  \definecolor{darkslateblue}{rgb}{0.28, 0.24, 0.55}
  \definecolor{slateblue}{rgb}{0.42, 0.35, 0.80}
  \definecolor{mediumslateblue}{rgb}{0.48, 0.41, 0.93}
  \definecolor{lightslateblue}{rgb}{0.52, 0.44, 1.00}
  \definecolor{mediumblue}{rgb}{0.00, 0.00,0.80}
  \definecolor{royalblue}{rgb}{0.25, 0.41, 0.88}
  \definecolor{blue}{rgb}{0.00, 0.00, 1.00}
  \definecolor{dodgerblue}{rgb}{0.12, 0.56, 1.00}
  \definecolor{deepskyblue}{rgb}{0.00, 0.75, 1.00}
  \definecolor{skyblue}{rgb}{0.53, 0.81, 0.92}
  \definecolor{lightskyblue}{rgb}{0.53, 0.81, 0.98}
  \definecolor{steelblue}{rgb}{0.27, 0.51, 0.71}
  \definecolor{lightsteelblue}{rgb}{0.69, 0.77, 0.87}
  \definecolor{lightblue}{rgb}{0.68, 0.85, 0.90}
  \definecolor{powderblue}{rgb}{0.69, 0.88, 0.90}
  \definecolor{paleturquoise}{rgb}{0.69, 0.93, 0.93}
  \definecolor{darkturquoise}{rgb}{0.00, 0.81, 0.82}
  \definecolor{mediumturquoise}{rgb}{0.28, 0.82, 0.80}
  \definecolor{turquoise}{rgb}{0.25, 0.88, 0.82}

  \definecolor{Turquoise}{rgb}{0.25, 0.88, 0.82}

  \definecolor{cyan}{rgb}{0.00, 1.00, 1.00}
  \definecolor{lightcyan}{rgb}{0.88, 1.00, 1.00}
  \definecolor{cadetblue}{rgb}{0.37, 0.62, 0.63}
  \definecolor{mediumaquamarine}{rgb}{0.40, 0.80, 0.67}
  \definecolor{aquamarine}{rgb}{0.50, 1.00, 0.83}
  \definecolor{darkgreen}{rgb}{0.00, 0.39, 0.00}
  \definecolor{darkolivegreen}{rgb}{0.33, 0.42, 0.18}
  \definecolor{darkseagreen}{rgb}{0.56, 0.74, 0.56}
  \definecolor{seagreen}{rgb}{0.18, 0.55, 0.34}
  \definecolor{mediumseagreen}{rgb}{0.24, 0.70, 0.44}
  \definecolor{lightseagreen}{rgb}{0.13, 0.70, 0.67}
  \definecolor{palegreen}{rgb}{0.60, 0.98, 0.60}
  \definecolor{springgreen}{rgb}{0.00, 1.00, 0.50}
  \definecolor{lawngreen}{rgb}{0.49, 0.99, 0.00}
  \definecolor{green}{rgb}{0.00, 1.00, 0.00}
  \definecolor{chartreuse}{rgb}{0.50, 1.00, 0.00}
  \definecolor{mediumspringgreen}{rgb}{0.00, 0.98, 0.60}
  \definecolor{greenyellow}{rgb}{0.68, 1.00, 0.18}
  \definecolor{limegreen}{rgb}{0.20, 0.80, 0.20}
  \definecolor{yellowgreen}{rgb}{0.60, 0.80, 0.20}
  \definecolor{forestgreen}{rgb}{0.13, 0.55, 0.13}
  
  \definecolor{ForestGreen}{rgb}{0.13, 0.55, 0.13}
  
  \definecolor{olivedrab}{rgb}{0.42, 0.56, 0.14}
  \definecolor{darkkhaki}{rgb}{0.74, 0.72, 0.42}
  \definecolor{khaki}{rgb}{0.94, 0.90, 0.55}
  \definecolor{palegoldenrod}{rgb}{0.93, 0.91, 0.67}
  \definecolor{lightgoldenrodyellow}{rgb}{0.98, 0.98, 0.82}
  \definecolor{lightyellow}{rgb}{1.00, 1.00, 0.88}
  \definecolor{yellow}{rgb}{1.00, 1.00 ,0.00}
  \definecolor{gold}{rgb}{1.00, 0.84, 0.00}
  \definecolor{lightgoldenrod}{rgb}{0.93, 0.87, 0.51}
  \definecolor{goldenrod}{rgb}{0.85, 0.65, 0.13}
  \definecolor{darkgoldenrod}{rgb}{0.72, 0.53, 0.04}
  \definecolor{rosybrown}{rgb}{0.74, 0.56, 0.56}
  \definecolor{indianred}{rgb}{0.80, 0.36, 0.36}
  \definecolor{saddlebrown}{rgb}{0.55, 0.27, 0.07}
  \definecolor{sienna}{rgb}{0.63, 0.32, 0.18}
  \definecolor{peru}{rgb}{0.80, 0.52, 0.25}
  \definecolor{burlywood}{rgb}{0.87, 0.72, 0.53}
  \definecolor{beige}{rgb}{0.96, 0.96, 0.86}
  \definecolor{wheat}{rgb}{0.96, 0.87, 0.70}
  \definecolor{sandybrown}{rgb}{0.96, 0.64, 0.38}
  \definecolor{tan}{rgb}{0.82, 0.71, 0.55}
  \definecolor{chocolate}{rgb}{0.82, 0.41, 0.12}
  \definecolor{firebrick}{rgb}{0.70, 0.13, 0.13}
  \definecolor{brown}{rgb}{0.65, 0.16, 0.16}
  \definecolor{darksalmon}{rgb}{0.91, 0.59, 0.48}
  \definecolor{salmon}{rgb}{0.98, 0.50, 0.45}
  \definecolor{lightsalmon}{rgb}{1.00, 0.63, 0.48}
  \definecolor{orange}{rgb}{1.00, 0.65, 0.00}
  \definecolor{darkorange}{rgb}{1.00, 0.55, 0.00}
  \definecolor{coral}{rgb}{1.00, 0.50, 0.31}
  \definecolor{lightcoral}{rgb}{0.94, 0.50, 0.50}
  \definecolor{tomato}{rgb}{1.00, 0.39, 0.28}
  \definecolor{orangered}{rgb}{1.00, 0.27, 0.00}
  \definecolor{red}{rgb}{1.00, 0.00, 0.00}
  \definecolor{hotpink}{rgb}{1.00, 0.41, 0.71}
  \definecolor{deeppink}{rgb}{1.00, 0.08, 0.58}
  \definecolor{pink}{rgb}{1.00, 0.75, 0.80}
  \definecolor{lightpink}{rgb}{1.00, 0.71, 0.76}
  \definecolor{palevioletred}{rgb}{0.86, 0.44, 0.58}
  \definecolor{maroon}{rgb}{0.69, 0.19, 0.38}
  \definecolor{mediumvioletred}{rgb}{0.78, 0.08, 0.52}
  \definecolor{violetred}{rgb}{0.82, 0.13, 0.56}
  \definecolor{magenta}{rgb}{1.00, 0.00, 1.00}
  \definecolor{violet}{rgb}{0.93, 0.51, 0.93}
  \definecolor{plum}{rgb}{0.87, 0.63, 0.87}
  \definecolor{orchid}{rgb}{0.85,0.44,0.84}
  \definecolor{mediumorchid}{rgb}{0.73,0.33,0.83}
  \definecolor{darkorchid}{rgb}{0.60,0.20,0.80}
  \definecolor{darkviolet}{rgb}{0.58,0.00,0.83}
  \definecolor{blueviolet}{rgb}{0.54,0.17,0.89}
  \definecolor{purple}{rgb}{0.63,0.13,0.94}
  \definecolor{mediumpurple}{rgb}{0.58,0.44,0.86}
  \definecolor{thistle}{rgb}{0.85,0.75,0.85}
\definecolor{snow}{rgb}{1.00,0.98,0.98}
\definecolor{ghostwhite}{rgb}{0.97,0.97,1.00}
\definecolor{whitesmoke}{rgb}{0.96,0.96,0.96} 
\definecolor{gainsboro}{rgb}{0.86, 0.86, 0.86}
\definecolor{floralwhite}{rgb}{1.00, 0.98, 0.94}
\definecolor{oldlace}{rgb}{0.99, 0.96, 0.90}
\definecolor{mistyroseen}{rgb}{0.98, 0.94, 0.90}
\definecolor{antiquewhite}{rgb}{0.98, 0.92, 0.84}
\definecolor{papayawhip}{rgb}{1.00, 0.94, 0.84}
\definecolor{blanchedalmond}{rgb}{1.00, 0.92, 0.80}
\definecolor{bisque}{rgb}{1.00, 0.89, 0.77}
\definecolor{peachpuff}{rgb}{1.00, 0.85, 0.73}
\definecolor{navajowhite}{rgb}{1.00, 0.87, 0.68}
\definecolor{moccasin}{rgb}{1.00, 0.89, 0.71}
\definecolor{cornsilk}{rgb}{1.00, 0.97, 0.86}
\definecolor{ivory}{rgb}{1.00, 1.00, 0.94}
\definecolor{lemonchiffon}{rgb}{1.00, 0.98, 0.80}

\usepackage{epsfig}
\newcommand{\fig}[1]{\parbox{1.5cm}{\epsfig{file=#1.eps,width=1.5cm}}}

\usepackage{simpler-wick}

\usepackage{graphicx,framed} % Include figure files
\usepackage{dcolumn} % Align table columns on decimal point
\usepackage{bm} % bold math
\usepackage{braket}
\usepackage{dsfont}
\usepackage{color,xcolor}
\usepackage{amsthm}
\usepackage{hyperref} % add hypertext capabilities
\usepackage[normalem]{ulem} %to strike the words
\usepackage{qcircuit}       % professional-quality tables
\usepackage[export]{adjustbox}

\usepackage{amsthm}

\usepackage{comment}

\usepackage{algorithm}
\usepackage[noend]{algpseudocode}
\algdef{SE}[DOWHILE]{Do}{doWhile}{\algorithmicdo}[1]{\algorithmicwhile\ #1}
\makeatletter
\def\algbackskip{\hskip-\ALG@thistlm}
\makeatother

\definecolor{lightblue}{RGB}{73,151,208}
\definecolor{crimson}{RGB}{140,41,53}

\hypersetup{
    colorlinks,
    linkcolor={crimson},
    citecolor={lightblue},
    urlcolor={lightblue}
}

\theoremstyle{definition}

\begin{document}

%{\footnotesize \hfill \blue{\today}}
%\vspace{5mm}

%\preprint{}

\title{The Casimir effect in wetting layers}

\author{Alessio~Squarcini}
\email{alessio.squarcini@uibk.ac.at}
\affiliation{Institut f\"ur Theoretische Physik, Universit\"at Innsbruck, Technikerstra{\ss}e~21A, A-6020 Innsbruck, Austria}

\author{Jos\'e M. Romero-Enrique}
\address{Departamento de F\'{\i}sica At\'omica, Molecular y Nuclear, \'Area de F\'{\i}sica Te\'orica, Universidad de Sevilla,
Avenida de Reina Mercedes s/n, 41012 Seville, Spain}
\address{Instituto Carlos I de F\'isica Te\'orica y Computacional, Campus Universitario Fuentenueva, Granada, Spain}

\author{Andrew O. Parry}
\address{Department of Mathematics, Imperial College London, London SW7 2AZ, United Kingdom}

\begin{abstract}
For a long time, the study of thermal effects at three-dimensional (3D) short-ranged wetting transitions considered only the effect of interfacial fluctuations. We show that an entropic Casimir contribution, missed in previous treatments, produces significant effects when it is included; in particular, mean-field predictions are no longer obtained when interfacial fluctuations are ignored. The Casimir term arises from the many different microscopic configurations that correspond to a given interfacial one. By employing a coarse-graining procedure, starting from a microscopic Landau-Ginzburg-Wilson Hamiltonian, we identify the interfacial model for 3D wetting and the exact form of the Casimir term. The Casimir contribution does not alter the Nakanishi-Fisher surface phase diagram; it significantly increases the adsorption near a first-order wetting transition and completely changes the predicted critical singularities of tricritical wetting, including the nonuniversality occurring in 3D arising from interfacial fluctuations. We illustrate how the Casimir term leads to a reappraisal of the critical singularities at wetting transitions.
\end{abstract}

\maketitle
%\tableofcontents

%====================================================================================================
%====================================================================================================
\section{Introduction}
The physics of interfaces between solids and fluids is a vast and complex field of study. Effective models, motivated by mesoscopic principles, can be used to gain physical insight. Examples of such models include the capillary-wave model of interfacial wandering \cite{buff} and models of surface growth \cite{KPZ}. However, for critical wetting in 3D systems with short-ranged forces, it is crucial to understand the details of the interfacial model and how it emerges from a microscopic framework. The phenomenon of critical wetting is characterized by the continuous growth of a liquid phase (for example) at a solid-gas interface (wall) as the temperature is increased towards a wetting temperature; for comprehensive reviews see, e.g., Refs.~\cite{dietrich_wetting_1988, Schick, FLN, BEIMR}. This process is associated with the divergence of a parallel correlation length, $\xi_\parallel$, which is characterized by an exponent $\nu_\parallel$. The theoretical framework for the description of interfaces interacting with substrates relies on capillary wave interfacial Hamiltonians incorporating the surface tension (or stiffness) and a binding potential determined by integrating the intermolecular forces over the volume of liquid. The 3D case with short-ranged forces turns out to be particularly intriguing because the binding potential itself arises from density fluctuations and decays on the scale of the bulk correlation length. Moreover, the space dimension $d=3$ coincides with the upper critical dimension and the original renormalization group (RG) studies predicted strong non-universal critical singularities \cite{LKZ_1983,BHL_1983,FH_1985} implying that $\nu_\parallel \approx 3.7$ for Ising-like systems, a value that is very different to the mean-field (MF) prediction \cite{PRBRE_2008,SREP_2022}  $\nu_\parallel=1$. This value for $\nu_{\parallel}$ has never been observed in experiments \cite{BONN_NATURE, BONN_PRL} nor very accurate Ising model simulations \cite{BLK_1986,BL_1988,PEB_1991, BLW_1989, BB_2013, BT_2021} which point towards an effective exponent $\nu_{\parallel}^{{\rm eff}} = 1.8 \pm 0.1$. To make the story even more interesting is the numerous series of analytical efforts to tackle the problem within a consistent theory. An initial attempt based on the possibility of a position dependence to the stiffness aggravated the issue, since it drives the wetting transition to the first-order \cite{FJ_1992, JF_1993, Boulter_1997}. This is in contrast to the findings of the simulations, as well as the predictions of the Nakanishi-Fisher global surface phase diagrams \cite{NF}, which posit a consistent relationship between wetting and surface criticality. More recent studies shown that the binding potential arising from correlations within the wetting layer is, in general, a non-local functional \cite{PREL_2004, PRBRE_2008_prl, PRBRE_2006, BPRRE_2009, RESPG_2018}. The binding potential can be expressed using a compact diagrammatic formulation, which could be used for walls and interfaces of arbitrary shape. This would remove the possibility that the wetting transition is driven first-order, restoring the global phase diagram.
\newline\indent
% REMOVED REFs: gennes_wetting_1985, PRBRE_2007, PREBR_2008
Notwithstanding the above achievements, the role played by bulk fluctuations has not been addressed. Since wetting occurs below the critical temperature, it has been tacitly assumed that bulk-like fluctuations are unimportant. This is equivalent to assume that non-classical exponents are the consequence of thermal wandering of the interface only. The derivation of interfacial Hamiltonians from more microscopic Landau-Ginzburg-Wilson (LGW) models renders explicit the above assumption. The constrained minimization is equivalent to a MF approximation of the trace over microscopic degrees of freedom. As a result of this assumption, the binding potential does not contain any information on an entropic contribution arising from the multiplicity of microscopic configurations that correspond to a given interfacial one -- a feature which is known to be important in molecular descriptions of free interfaces \cite{Pedro_1, Pedro_2, Pedro_3, MCT}. Effects of entropic repulsion at the level of one- and two-point correlation functions for both planar \cite{ST_droplet_long} and wedge-shaped boundaries \cite{DS_wedgebubble} have been investigated within the exact theory of phase separation in $d=2$ \cite{ST_threepoint, ST_fourpoint}.
%REMOVED REFs: DV, Squarcini_Multipoint, ST_droplet_MC, Squarcini_Tinti_SciPots, DS_wedge
The entropic contribution to the binding potential is equivalent to the thermal Casimir effect emerging in a system close to a second order phase transition point experiencing a geometrical restriction, an effect predicted in 1978 by Fisher and de Gennes \cite{FdG}. Sometimes, this is also stated in different words by saying that the effect arises as a restriction of bulk fluctuations in a confined fluid \cite{BCN, ES, HGDB}. The thermal Casimir effect has been the subject of intense studies \cite{MD_2018, GambassiDietrich2024}. In general, right at bulk criticality the thermal Casimir force is long-ranged but it is always present \footnote{See, e.g., Ref.~\cite{Squarcini_Casimir} and references therein for results in conformal field theory.}, even away from the critical point where it decays on the scale of the bulk correlation length \cite{AM2010, DS2015}. A distinctive feature of wetting with short-range forces is that the additional entropic or low-temperature Casimir term in the binding potential displays a comparable range to the MF contribution, as recently evidenced \cite{SREP_2022}. Recent findings \cite{SREP_2022} have demonstrated that, in the context of critical wetting, the Casimir term gives rise to an exceptionally narrow asymptotic regime for the growth of $\xi_{\parallel}$. This latter phenomenon is characterized by an effective exponent that aligns with the predictions of the Ising model, thus resolving a long-standing controversy in the field.
\newline\indent
This paper presents recent advances in our comprehension of the implications of the Casimir effect on critical singularities at wetting transitions. The conclusions of this paper are largely contingent upon the diagrammatic formalism of the Casimir contribution to the binding potential, which was originally derived \cite{SREP_2022} through the proper application of the constrained trace to the LGW model. This approach allows for the exact determination of the interfacial Hamiltonian for 3D wetting. Following a review of the principal findings concerning the diagrammatic formalism, we will proceed to examine the characteristics of the Casimir potential in 3D and its implications for wetting in higher dimensions. In particular, we will examine which characteristics of wetting remain unaltered and which are affected by the introduction of a Casimir term for a flat wetting layer of uniform thickness. Finally, we will briefly discuss the influence of curvature by presenting recent findings \cite{SREP_in_preparation} on the Casimir potential in a system of concentric spheres and cylinders.
\newline\indent
%\red{In general, this can be described accurately using a simple interfacial model incorporating the surface tension (or stiffness) and a bind-potential determined by integrating the intermolecular forces over the volume of liquid.}
%Greater care is required in 3D with short-ranged forces where the binding potential itself arises from density fluctuations and decays on the scale of the bulk correlation length. 
%Exact results for the planar Ising model \cite{Abraham_review, Abraham_1980} and from quantum field theory \cite{Squarcini_Tinti}. Classical scenario \cite{EbnerSaam_1977, cahn_critical_1977}. Brezin-Halperin-Leibler \cite{BHL_1983}, Fisher-Jin \cite{FJ_1992}. Nonlocal model \cite{PREL_2004, PRBRE_2006}.

%====================================================================================================
%====================================================================================================
\section{The model}
We begin by recalling the model, which is formulated in terms of a LGW Hamiltonian based on a magnetization-like order-parameter \cite{NF}
\begin{equation}
H[m]= \int d\mathbf{r} \left(\frac{1}{2} (\boldsymbol\nabla m)^2 +\phi(m)\right) + \int_{\mathcal{S}_\psi} d\mathbf{s}  \phi_1 (m(\mathbf{s})) \, ,
\label{HLGW}
\end{equation}
where $\phi(m)$ is a double well potential (with Ising symmetry) $m_0$ denotes the spontaneous magnetization and $\kappa$ the inverse bulk correlation length. The surface potential is $\phi_1= c(m-m_s)^2/2$, with $c$ the enhancement parameter and $m_s$ the favored order-parameter at the wall $\mathcal{S}_\psi$ with Monge parameterization $(\mathbf{x}, \psi)$. Equivalently, $h_1 = c m_s$ is the surface field. The MF surface phase diagram is determined by minimizing the functional $H[m]$. We recall that for a planar wall the phase diagram exhibits critical wetting (for $c>\kappa$) and first-order wetting transitions (for $c<\kappa$). Consider now the interface separating the gas and the liquid phases and let $\ell$ be its height with respect to the wall. It is possible to derive an interfacial model $H_I[\ell]$ with the interfacial co-ordinate determined by a crossing criterion so that $m(\mathbf{x}, \ell(x))=0$ on the interface $\mathcal{S}_\ell$. Formally, this is identified via $\exp(-\beta H_I[\ell])=\int \mathcal{D}'m \exp (-\beta H[m])$ where $\beta=1/k_BT$ and the prime denotes a constrained trace over microscopic degrees of freedom respecting the crossing criterion, as shown by Fisher and Jin \cite{FJ_1991}. This procedure yields
%REMOVED: ; in the following, we will consider $c>0$. 
\begin{equation}
\label{ }
H_I[\ell]=\gamma A_{l}+W[\ell,\psi] \, ,
\end{equation}
where the first term is the surface tension times the interfacial area describing the free interface (ignoring curvature terms) and $W[\ell,\psi]$ is the binding potential functional describing the interaction with the wall. The customary way to evaluate the constrained trace is to ignore bulk fluctuations. This amounts to considering a MF approximation that identifies $H_I[\ell]=H[m_\Xi]$ where $m_\Xi$ is the \emph{unique} profile that minimizes the LGW Hamiltonian subject to the crossing criterion. A substantial simplification occurs by employing the double parabola (DP) approximation $\phi(m)=\kappa^2(|m|- m_0)^2/2$. The quadratic nature of the double well allows for analytical progress that, however, do not spoil the essential physical features. Indeed, it has been shown how to retrieve in a perturbative fashion the full $m^{4}$ model starting from the DP one \cite{PRBRE_2007}. For the sake of simplicity, we can consider a flat interface of thickness $\ell$ on a planar wall ($\psi=0$) of lateral  area $L_\parallel^2$. The binding potential functional reduces to the binding potential function $w_{MF}=W_{MF}/ L_\parallel^2$ which has the well-known exponential expansion \cite{FJ_1991}
\begin{equation}
\frac{w_{MF}(\ell)}{\gamma}\approx - \frac{2t  c }{c+\kappa} \textrm{e}^{-\kappa \ell} +\frac{c-\kappa}{c+\kappa} \textrm{e}^{-2\kappa \ell} \, ,
\label{wflatmf}
\end{equation}
where $\gamma=\kappa m_0^2$ is the surface tension and $t=(m_0-m_s)/m_0$ is the temperature-like scaling field for critical wetting. It has to be noticed that the first term is proportional to $T-T_{\rm w}^{\rm MF}$; hence, it changes sign at the MF wetting temperature. The minimum of $w_{MF}(\ell)$ determines the MF thickness of the wetting layer, while the curvature of $w_{MF}(\ell)$ at the minimum determines $\xi_\parallel$. Both these length scales diverge continuously as $t\to 0$ when $c>\kappa$. It should be noted that for tricritical wetting ($c=\kappa$) and first-order wetting ($c<\kappa$) it is necessary to include the next-order decaying exponential term. For non-planar interfaces (and walls) the non-local MF functional $W_{MF}[\ell,\psi]$ can also be determined exactly using boundary integral methods based on the Green function in the wetting layer \cite{PRBRE_2006,RESPG_2018}. Within the DP approximation it is possible to evaluate the constrained trace in closed form; this yields the exact binding potential
\begin{equation}
\label{08122024_1727}
W[\ell,\psi]=W_{MF}[\ell,\psi] + W_{C}[\ell,\psi] \, ,
\end{equation}
which contains a Casimir correction on top of the MF result. The proof of the ``one-loop'' functional for the Casimir contribution given in \eqref{08122024_1727} is rather technical. A sketch of the derivation for an arbitrarily shaped interface and wall is outlined in Ref.~\cite{SREP_2022}. The proof for a $d$-dimensional slab is presented in Ref.~\cite{SREP2023} by using a continuum transfer-matrix (path-integral) method. In the discussion that follows, we will limit ourselves to the final result for $W_C[\ell,\psi]$ and the implications for wetting transitions of all orders. The Casimir contribution admits a diagrammatic representation in terms of wetting diagrams similar to the terms in the MF contribution; however, the diagrammatic structure exhibits a distinct topology. We introduce two kernels that connect positions, with respective transverse coordinates $\textbf{s}$, $\textbf{s}^{\prime}$ (denoted by the open circles) on the interface (upper wavy line) and wall (lower wavy line). The building blocks of the diagrammatic expansion are the following diagrams:
\begin{equation}
\fig{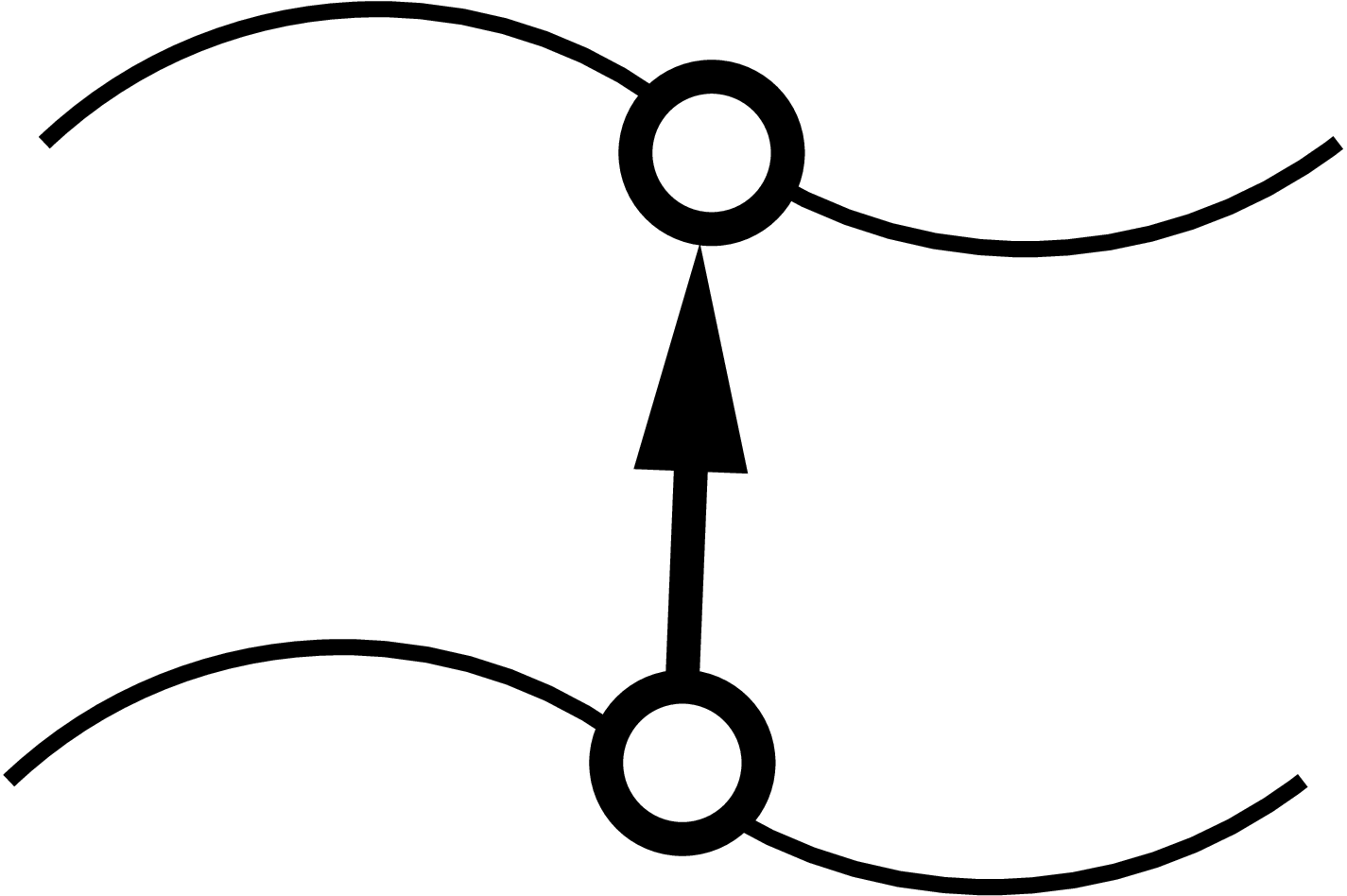} =  \mathbf{n}(\mathbf{s}) \mathbf{\cdot} \frac{\mathbf{s}^{\prime}-\mathbf{s}}{|\mathbf{s}-\mathbf{s}^{\prime}|^2}\left(1+ \frac{1}{\kappa|\mathbf{s}-\mathbf{s}^{\prime}|}\right)\textrm{e}^{-\kappa|\mathbf{s}-\mathbf{s}^{\prime}|} \, ,
\label{defdiagrams3bisbis}
\end{equation}
which was introduced \cite{RESPG_2018} in the derivation of $W_{MF}[\ell,\psi]$. Here $\textbf{n}(\textbf{s})$ is the normal at the wall. We also define
\begin{eqnarray}
\fig{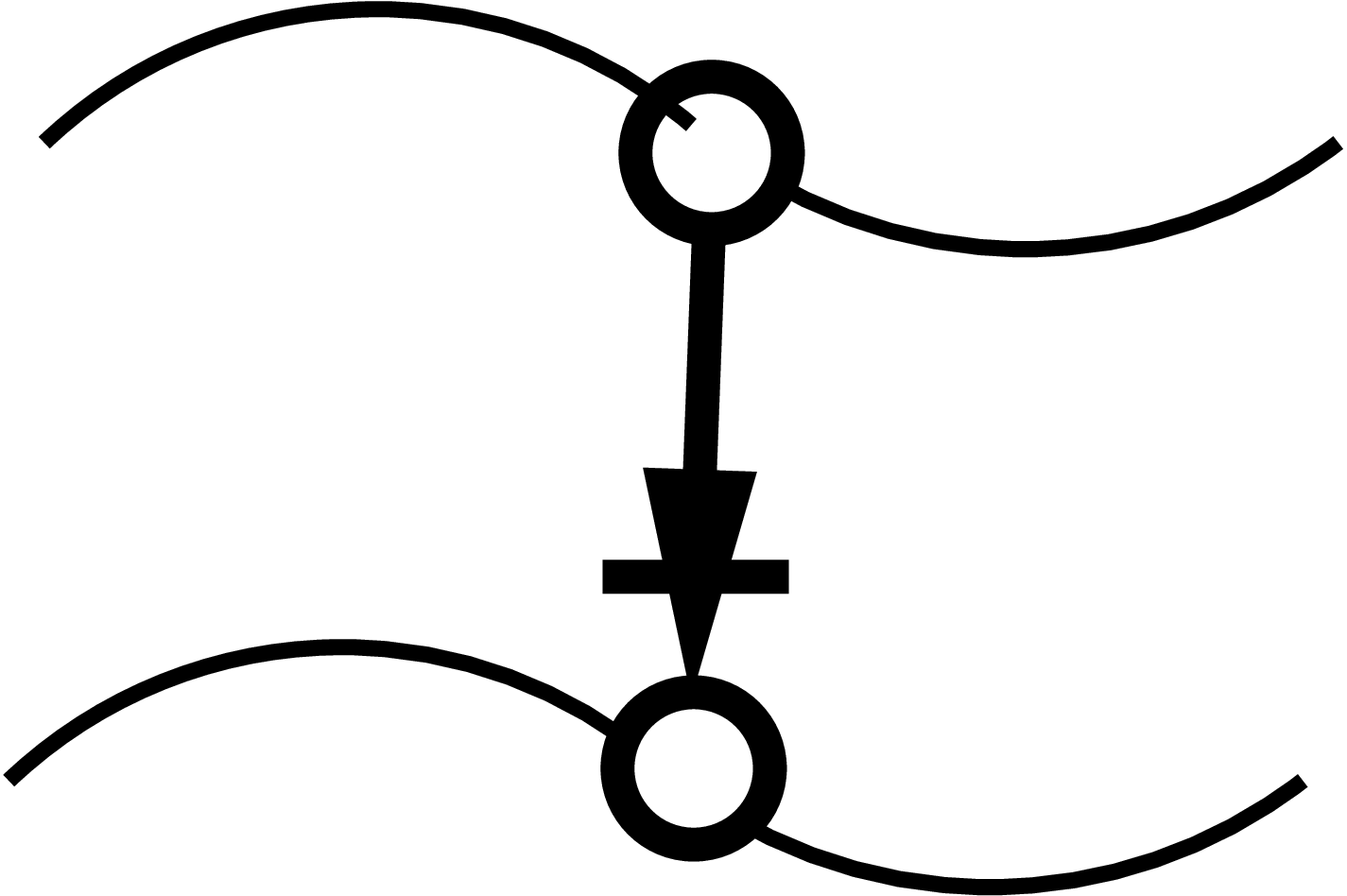} 
= \frac{1}{2\pi} \int_0^\infty dq\ q \frac{g+\kappa_q}{g-\kappa_q} J_0(q\rho)\exp(-\kappa_q \ell) \, ,
\label{defdiagrams2bisbis}
\end{eqnarray}
where $\rho$ and $\ell$ are, respectively, the transverse and normal coordinates of $\mathbf{s}-\mathbf{s}'$, $J_0(z)$ is the Bessel function of the first kind and zero order and $\kappa_q\equiv \sqrt{\kappa^2+q^2}$. 
In terms of the above diagrams, the Casimir term for a wetting film at a wall of arbitrary shape can be written as follows
\begin{equation}
\beta W_{C}[\ell,\psi]=\frac{1}{2} \fig{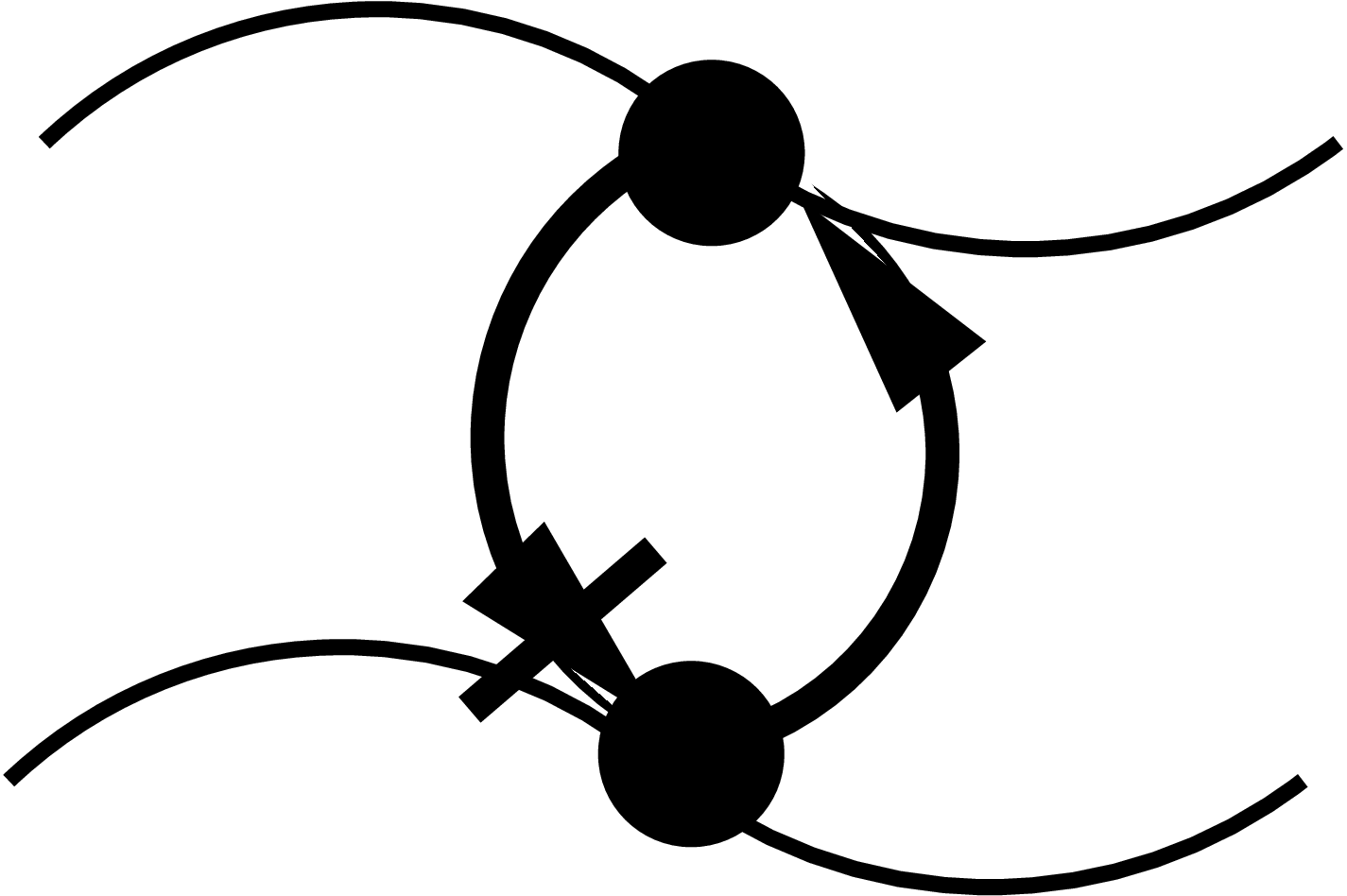} - \frac{1}{4} \fig{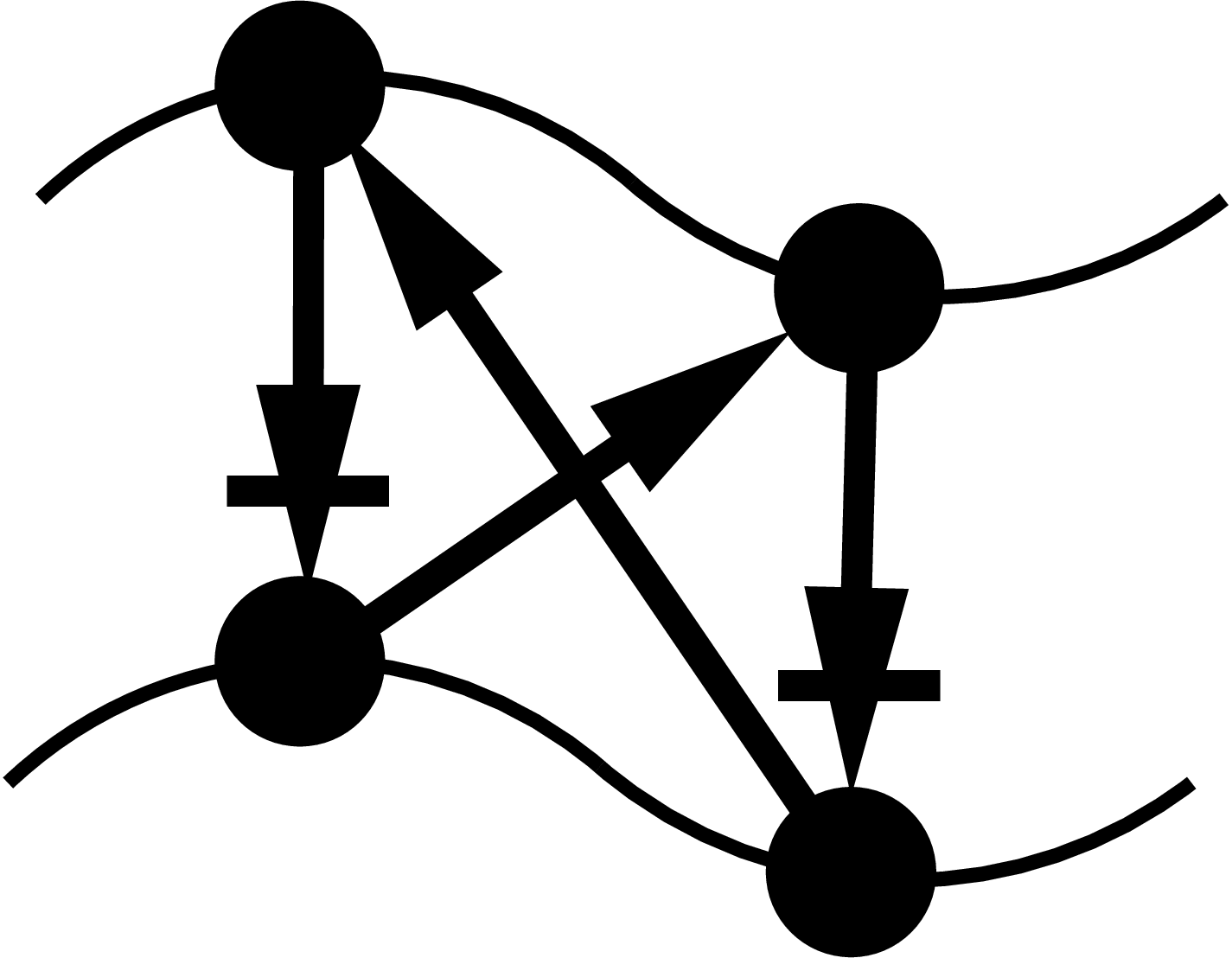}+\ldots \, .
\label{wcasimirdiagrammatic}
\end{equation}
Here the black dots simply imply integration over the points on the wall and interface with the appropriate measure for the local area. For thick wetting films, only the first term, which we refer to as $(\Omega_C)_1^1$ is required.

%====================================================================================================
%====================================================================================================
\section{Discussion of results}
We consider a uniform wetting layer on a planar wall and check which MF predictions are altered by the Casimir term $w_C(\ell)=W_C/L_\parallel^2$ to the binding potential.
\newline\indent
%====================================================================================================
%====================================================================================================
\subsection{Properties of the Casimir potential}
\begin{figure}[t!]
\includegraphics[width=90mm]{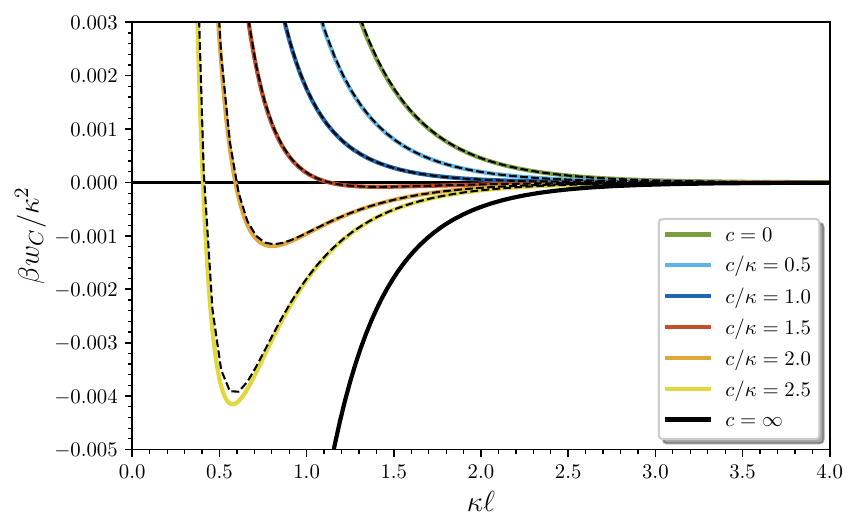}
\caption{
The Casimir contribution to the binding potential for a wetting layer of uniform thickness. The qualitative change from attraction to repulsion near the MF tricritical point is shown for the surface enhancements in the inset. The dotted lines show the comparisons between the full result, Eq. (\ref{wflat}), and the leading term arising from $(\Omega_C)_1^1$, which is near exact.
\label{fig6}}
\end{figure}

The Casimir potential in $d=3$ takes the form
\begin{equation}
\beta w_{C}(\ell)= \frac{1}{4\pi}\int dq\ q\ln\left(1-\frac{c-\kappa_q}{c+\kappa_q} \textrm{e}^{-2\kappa_q\ell}\right) \, .
\label{wflat}
\end{equation}
This result is similar in form to $w_{MF}(\ell)$ but controlled by $c$ rather than $t$. We observe that for $\kappa\ell \ll 1$, $w_{C}(\ell) \propto 1/\ell^{2}$, which is the familiar long-ranged Casimir limit \cite{FdG}. For the sake of completeness, we mention that in $d$-dimensions the corresponding result is $\beta w_{C}(\ell) \propto \ell^{-(d-1)}$, a standard result for the thermodynamics Casimir effect, although the DP approximation applies for thick wetting films, i.e., $\kappa\ell \gg1$. In general, for $c>\kappa$ the potential is repulsive at short distances, attractive at large distances, and possesses a minimum that diverges continuously as $c$ approaches $\kappa$. On the other hand, for $c <\kappa$, the potential is purely repulsive, as illustrated in Fig.~\ref{fig6}.
A remarkable fact is the occurrence of a qualitative change in the vicinity of the MF tricritical point, $c \approx \kappa$, where the Casimir potential
behaves as
\begin{equation}
\beta w_{C}(\ell)\approx \frac{\textrm{e}^{-2\kappa \ell}}{32\pi\ell^2}(1+2(\kappa-c)\ell) \, ,
\label{asymptomegac}
\end{equation}
meaning that the force is repulsive and that the average thickness diverges as $\ell_{\min} \sim [2(c-\kappa)]^{-1}$. The nearly accurate expression \eqref{asymptomegac} can be obtained at the level of the leading-order diagrams by using the approximation $\ln (1+x) \approx x$ in (\ref{wflat}), which is legitimate for $\kappa\ell\gg1$. The corresponding integral can be carried out exactly, and the result can be expressed in terms of the incomplete gamma function. Diagrams of order $\exp(-4\kappa\ell)$ -- ignored within this approximation -- can be safely neglected for the study of fluctuation effects. Another limiting case is provided by Dirichlet boundary conditions; this can be retrieved in the limit $c \rightarrow \infty$ and the result is $\beta w_{C}(\ell) = -(1+2\kappa\ell) \textrm{e}^{-2\kappa\ell}/\pi\ell^{2}$. This regime corresponds to critical wetting with fixed surface magnetization. Quite remarkably, when $c \rightarrow 0$ we find the very same result up to an overall different sign; however, this duality persists only at the level of diagrams belonging to $(\Omega_{1}^{2})^{C}$. We conclude with some observations about results in higher dimensions. For arbitrary dimension $d$ and $c \neq \kappa$ the Casimir potential decays at large distances as $\pm \ell^{-(d-1)/2}\textrm{e}^{-2\kappa \ell}$. The overall sign is $+$ for $c > \kappa$, hence the force is attractive, while the overall sign is $-$ for $c < \kappa$. However, exactly at MF tricriticality, $c = \kappa$, the large-distance decay is governed by the repulsive term $\ell^{-(d+1)/2}\textrm{e}^{-2\kappa \ell}$.% It is a general fact that $w_{C}(\ell)$ switches from attraction to repulsion at $c = \kappa$.

%====================================================================================================
%====================================================================================================
\subsection{Repercussions due to the Casimir term}
The presence of a previously missing Casimir term suggests that the study of fluctuation effects on wetting layers should be completely revisited. A careful analysis indeed confirms that this is the case. In particular, we may ask: are the MF critical singularities affected by the bulk-like thermal effects? Does the Casimir potential affect the Nakanishi-Fisher phase diagram? While some aspects of wetting are unaltered, others are changed drastically, even before we consider the role of interfacial undulations. Indeed, thermal effects due to the Casimir term yield significant effects even when the interface is flat.
\newline\indent
An important result is that fluctuations do not alter the general features of the phase diagram \cite{SREP_2022}, such as its topology, a feature in agreement with simulations. As a result, the Nakanishi-Fisher surface phase diagram is unaffected qualitatively so that, for example, critical wetting still occurs for $c>\kappa$ as $t \to 0$ with $\xi_\parallel\sim t^{-\nu_{\parallel}}$ and $\nu_{\parallel}=1$, as before. Nonetheless, tricritical wetting transitions are dramatically different \cite{SREP_2022} since it is the Casimir term that determines the repulsion. Indeed, at wetting tricriticality the prefactor of the term $\propto \exp(-2\kappa \ell)$ vanishes (see \eqref{wflatmf}) and the MF binding potential contributes with a term of $O(\exp(-3\kappa\ell))$, which is negligible with respect to the Casimir potential, which dominates the scene. In dimension $d>1$, this term decays as $\textrm{e}^{-2\kappa \ell}/\ell^{\frac{d+1}{2}}$ which, at finite $T$, {\it{always}} dominates over the higher-order MF contribution. In particular, in dimension $d>3$, where interfacial fluctuations are irrelevant, it follows that the parallel correlation length diverges as $\xi_\parallel \sim 1/ (t |\ln t| ^{\frac{d-1}{2}})$ in contrast to the strict MF prediction $\xi_\parallel \sim t^{-\frac{3}{4}}$ which misses the Casimir term. Then, another significant difference is the thickness of the wetting layer: MF predicts $\kappa\ell \approx 2\ln(1/t)$ while MF plus the Casimir term yields $\kappa\ell \approx \ln(1/t) - (d-1)\ln\ln(1/t)$.
\newline\indent
The Casimir term entails repercussions also for first-order wetting transitions \cite{SREP_2022}. The film thickness $\ell_{eq}$ at the transition, which, recall, smoothly increases as we follow the line of wetting transitions toward the tricritical point. MF theory predicts $\kappa \ell_{eq} \approx - \ln(1-c/\kappa)$ while, in $d=3$, the Casimir contribution alters this to
\begin{equation}
\ell_{eq}\approx  \bigl[ 16\pi \beta\gamma \left(1-c/\kappa \right) \bigr]^{-1/2} \, .
\end{equation}
The Casimir repulsion therefore dramatically increases the adsorption for weakly first-order transitions and similarly enhances $\xi_{\parallel}$. As we anticipated, the Casimir effect does not alter the topology of the surface phase diagram; however, it produces a shift of the field $t_{w}$ at which first-order wetting occurs \cite{SREP_in_preparation}. At MF level, and in the vicinity of tricriticality, $t_{w} \sim - (\kappa-c)^{2}$ for $c \rightarrow \kappa$. The introduction of the Casimir term alters the above result to $t_{w} \propto - (\kappa-c)^{3/2} \exp(-1/\sqrt{1-c/\kappa})$, which, for $c \rightarrow \kappa$, lies close to the line of critical wetting, $t=0$. This example is confirming once more the importance of the Casimir effect.
\newline\indent
Lastly, so far we have considered the effect of thermal fluctuations in the slab geometry. However, the formalism allows us to study interfaces of arbitrary shape, and therefore it would be interesting to study the interplay of thermal fluctuations and curvature effects. A very interesting case of study is provided by the configuration of concentric spheres and cylinders, for which it is possible to work out the Casimir potential in an exact fashion \cite{SREP_in_preparation}. Let us consider a spherical (or cylindrical) wall of radius $R$ (i.e., $\psi=R$) and a concentrical liquid film of thickness $\ell$, i.e., the liquid-vapor surface is located at a distance $R+\ell$ from the center. The Casimir potential per unit area takes the form
\begin{equation}
\label{08122024_2150}
\beta w_{C}^{\rm (s)}(\ell) = \beta w_{C}^{\rm (planar)}(\ell) - \mathcal{C}_{s} \frac{ 1+2\kappa\ell }{ 16\pi R\ell } \textrm{e}^{-2\kappa\ell} \, ,
\end{equation}
the superscript stands for the shape, either a sphere or cylinder. Here, $\mathcal{C}_{s}=1$ for concentric spheres and $\mathcal{C}_{c}=1/2$ for concentric cylinders. The derivation of the result \eqref{08122024_2150} is rather technical as it involves a calculation based on the zeta function regularization and will be presented elsewhere \cite{SREP_in_preparation}. The structure of the result invites speculation about the origin of the factor $\mathcal{C}_{s}$. It is natural to conjecture that $\mathcal{C}_{s}$ is related to the mean curvature of the surface, which is $H = (\kappa_{1} + \kappa_{2})/2$, where $\kappa_{j}$ are the principal curvatures. For a cylinder, $H_{c}=(2R)^{-1}$, while for a sphere, $H_{s}=1/R$. These features are compatible with $\mathcal{C}_{s} = R H_{s}$.

%\cite{DNF}

%====================================================================================================
%====================================================================================================
\section{Conclusions}
To summarize, previous studies of 3D short-ranged wetting have missed a thermal Casimir, or entropic, contribution to the binding potential. This entropic term is due to the multitude of different microscopic configurations that correspond to a given interfacial one. In this paper, we highlighted the most important repercussion of the Casimir term for wetting layers, recently computed \cite{SREP_2022} by using a using a boundary integral method \cite{RESPG_2018} which can be cast as a diagrammatic expansion. This method is actually equivalent to the Li-Kardar formalism \cite{LK_1992} for the description of medium-mediated fluctuation-induced forces between inclusions in a critical medium. We then illustrated which features of wetting are unchanged and which are altered by the Casimir effect even before considering interfacial fluctuations. In particular, by considering the slab geometry, we showed that the Casimir effect does not alter the Nakanishi-Fisher phase diagram, a finding in agreement with simulation studies. However, the scaling field $t_{w}$ controlling first-order wetting is significantly altered, especially close to MF tricriticality. Then, we showed that critical wetting is not altered by the Casimir effect. The most striking results are the dramatic consequences of the Casimir effect on the critical singularities of both tricritical and first-order wetting. In particular, the singularities at wetting tricriticality resemble those of MF critical wetting. Preliminary results show qualitative changes also at first-order wetting, for which a detailed investigation is currently in progress \cite{SREP_in_preparation}.
\newline\indent
Another remarkable result concerns the commonly accepted paradigm of MF theory. Our results show that the predictions of MF theory are not valid for dimension $d > d^{*}$ -- i.e., above the upper critical dimension $d^{*}$--, as have always been thought previously. This is because MF predictions happen to be altered by the thermal Casimir effect even for $d>3$ \cite{SREP_in_preparation}. As a matter of fact, MF is only recovered on setting $k_{B}T=0$ or in the limit $d \rightarrow \infty$, meaning that previous interpretations of the Ginzburg criterion for short-ranged tricritical wetting should be revisited too. Lastly, we commented on some recent results about curvature and thermal effects, a topic that is still untouched in the context of wetting phenomena. As recently shown \cite{SREP_2022}, the interplay of interfacial fluctuations due to capillary waves and thermal fluctuations stemming in the Casimir term allowed us to show that the exponent $\nu_{\parallel}$ for critical wetting is in quantitative agreement with Ising model simulations, resolving a longstanding controversy. Looking at perspectives, it would be extremely interesting to investigate nonlocality, capillary waves, and the Casimir effect at tricritical wetting. Finally, we mention that the entropic contribution is a general feature that occurs also in systems with short-ranged fluid-fluid but long-ranged wall-fluid forces \cite{ESW_2019, PM_2024}. In a nutshell, we showed how hard it is to overstate the importance of the Casimir effect  in the context of interfacial phenomena.
% These findings are essential to reassess the role of the upper critical dimension for tricritical wetting.
%We show that the Casimir term plays an important role at wetting transitions of all orders forcing a reappraisal of the accuracy of MF and subsequent RG theory and altering even the values of critical exponents.
%changes the interpretation of thermal fluctuation effects at wetting transitions which arise both from it and from capillary-wave-like interfacial wandering. Both are missing in MF descriptions. 
%tricritical wetting in higher dimensions
%\red{In summary, in this paper we have pointed out that previous theories of 3D short-ranged wetting have missed a thermal Casimir, or entropic, contribution to the binding potential, which we have determined exactly for the LGW model within the DP approximation. This decays exponentially, similar to the MF contribution and is qualitatively different for first-order and critical wetting.
% By using the numerical renormalization group we show that, for critical wetting, the asymptotic regime is extremely narrow with the growth of the parallel correlation length characterised by an effective exponent 
% The Casimir contribution is qualitatively different for first-order, critical and tri-critical wetting transitions and substantially alters previous predictions for critical singularities bringing them much closer to the simulation results.}
%Ref.~\refcite{autbk} that 
%Refs.~\cite{jpap,colla,autbk}, \refcite{rvo} and \refcite{pro}
%Ref.~\verb|\cite{name}|''.

%%%%%%%%%%%%%%%%%%%%%%%%%%%%%%%%%%%%%%%%%%%%%%%%%%%%%%%%%%%%
\section*{Acknowledgments}
AS gratefully acknowledges FWF Der Wissenschaftsfonds for funding through the Lise-Meitner Fellowship (Grant DOI 10.55776/M3300) and thanks with pleasure the organizers of the ``Fifth International Symposium on the Casimir Effect'' held in Piran (Slovenia) for the stimulating atmosphere. JMRE acknowledges financial support from the Ministerio de Ciencia, Innovacion y Universidades (Spain) through the grant PID2021-126348NB-I00.

%%%%%%%%%%%%%%%%%%%%%%%%%%%%%%%%%%%%%%%%%%%%%%%%%%%%%%%%%%%%
%\cleardoublepage
%\appendix
%%%%%%%%%%%%%%%%%%%%%%%%%%%%%%%%%%%%%%%%%%%%%%%%%%%%%%%%%%%%

%%%%%%%%%%%%%%%%%%%%%%%%%%%%%%%%%%%%%%%%%%%%%%%%%%%%%%%
\bibliography{bibliography}
\end{document}